\documentstyle[prd,aps,floats]{revtex}

\begin{document}
\preprint{SUSSEX-AST 98/3-5, gr-qc/9803094}
\draft

%
%

\input epsf
\renewcommand{\topfraction}{1.0}
\twocolumn[\hsize\textwidth\columnwidth\hsize\csname 
@twocolumnfalse\endcsname

\title{Brans--Dicke Boson Stars: Configurations and Stability through
Cosmic History}

\author{Diego F.~Torres${^{1,2}}$, Franz E.~Schunck${^1}$ and Andrew 
R.~Liddle${^1}$}

\address{${^1}$Astronomy Centre, University of Sussex, Brighton BN1 
9QJ, Great Britain}
\address{${^2}$Departamento de F\'{\i}sica,  Universidad Nacional 
de La Plata, C.C. 67, 1900 La Plata, Argentina}

\date{\today} 

\maketitle

\begin{abstract}

We make a detailed study of boson star configurations in Jordan--Brans--Dicke
theory, studying both equilibrium properties and stability, and considering
boson stars existing at different cosmic epochs.  We show that boson stars
can be stable at any time of cosmic history and that equilibrium stars are
denser in the past.  We analyze three different proposed mass functions for
boson star systems, and obtain results independently of the definition
adopted.  We study how the configurations depend on the value of the
Jordan--Brans--Dicke coupling constant, and the properties of the stars under
extreme values of the gravitational asymptotic constant.  This last point
allows us to extract conclusions about the stability behaviour concerning the
scalar field.  Finally, other dynamical variables of interest, like the
radius, are also calculated.  In this regard, it is shown that the radius
corresponding to the maximal boson star mass remains roughly the same during
cosmological evolution.

\end{abstract}

\pacs{PACS numbers: 04.40.Dg, 04.50.+h \hspace*{0.8cm} Sussex preprint 
SUSSEX-AST 98/3-5, gr-qc/9803094}

\vskip2pc]

\section{Introduction}

Although boson stars \cite{KAUP,RB,R1,R2} are so far entirely theoretical
constructs, they give rise to one of the simplest possible stellar
environments in which to study gravitational phenomena mathematically.  One
can find numerical solutions which are nonsingular and yet exhibit strong
gravitational effects.  Many of their properties bear close resemblance to
those of neutron stars.

Boson stars were first conceived as Klein--Gordon geons --- systems held
together by gravitational forces and composed of classical fields.  They are
a gravitationally-bound macroscopic state made up of scalar bosons.  As with
neutron stars, the pressure support which leads to their existence is
intrinsically quantum.  For neutron stars, the pressure support derives from
the Pauli exclusion principle, and for boson stars this is replaced by
Heisenberg's uncertainty principle.  Assuming that the quantum state contains
sufficient particles for gravitational effects to be important, and that
particle interactions can be neglected, an estimate of the mass is readily
obtained as follows.  For a quantum state confined into a region of radius
$R$, and with units given by $h=c=1$, the boson momentum is $p=1/R$.  If the
star is moderately relativistic, $p\simeq m$, then $R \simeq 1/m$.  If we
equate $R$ with the Schwarzschild radius $2M/m_{{\rm Pl}}^2$ (recall that $G
\equiv m_{{\rm Pl}}^{-2}$), we find $M \simeq m_{{\rm Pl}}^2/m$.

In practice, one assumes the existence of a classical scalar field with a
given Lagrangian density, and adopts an ansatz for its time dependence which
implicitly encodes the Heisenberg uncertainty.  This time dependence is of
course of a form which still permits a static metric.  With these
ingredients, one then solves Einstein's equations, something which must be
done numerically.  When no self-interaction term is present in the Lagrangian
density, the masses concur with the estimate above.  However, if
self-interaction is present it is typically the dominant contributor to
pressure support, and leads instead to masses of order $m_{{\rm Pl}}^3/m^2$.
If the boson mass is comparable to a nucleon mass, this order of magnitude is
comparable to the Chandrasekhar mass, about 1$M_\odot$ \cite{COLPI}.  Thus,
boson stars arise as possible candidates for non-baryonic dark matter, and
are possibly detectable by microlensing experiments.

Boson stars have been widely studied in general relativity, where the basic
model has been extended in various ways, such as including a $U(1)$ charge
\cite{4-boson}, allowing a mixture of boson and fermion components
\cite{8-boson}, or including a non-minimal coupling of the boson field to
gravity \cite{6-boson}.  These and others works are summarized in two reviews
\cite{R1,R2}, and more recently in Ref.~\cite{LAST-SCHUNCK}.  The possibility
of direct observational detection of boson stars was studied recently in
Ref.~\cite{LIDDLE-SCHUNCK}, where it was asked whether radiating baryonic
matter moving {\em within} a boson star could be converted into an
observational signal.  Unfortunately, any direct detection looks a long way
off.

Given the simplicity of the boson star, it is natural to examine boson star
solutions in theories of gravity other than general relativity, to
examine whether new phenomena arise.  The most-studied class of such theories
are the scalar--tensor theories of gravity \cite{12-boson}, which include the
Jordan--Brans--Dicke (JBD) theory as a special case.  In these, Newton's
gravitational constant is replaced by a field $\phi$ known as the
Brans--Dicke field, the strength of whose coupling to the metric is given by
a function $\omega(\phi)$.  If $\omega$ is a constant, this is the
JBD theory \cite{BD}, which is the simplest scenario
one may have in this framework.  General relativity is attained in the limit 
$1/\omega
\rightarrow 0$.  To ensure that the weak-field limit of this theory agrees
with current observations, $\omega$ must exceed 500 at 95\% confidence
\cite{12-boson} from solar system timing experiments, i.e.  experiments
taking place in the current cosmic time.  This limit is both stronger and
less model-dependent than limits from nucleosynthesis \cite{CASAS}.
Scalar--tensor theories have regained popularity through inflationary
scenarios based upon them \cite{13-boson}, and because a JBD model with
$\omega =-1$ is the low-energy limit of superstring theory \cite{14-boson}.

The first scalar--tensor models of boson stars were studied by Gunderson and
Jensen \cite{GUNDERSON}, who concentrated on JBD theory with $\omega=6$.
This was generalized by Torres \cite{TORRES_BOSON}, both to other JBD
couplings and to some particular scalar--tensor theories with non-constant
$\omega(\phi)$ chosen to match all current observational constraints.  This
allowed a study of some models which, inside the structure of the star, have
couplings deviating greatly from the large value required today.  The
conclusion is that boson star models can exist in any scalar--tensor gravity,
with masses which are always smaller than the general relativistic case (for
a given central scalar field density), irrespective of the coupling.

A vital point to consider is that when one finds cosmological solutions in
scalar--tensor theories, the gravitational coupling is normally evolving.
This has important implications for astrophysical objects, because it means
that the asymptotic boundary condition for the $\phi$ field is in general a
function of epoch.  One can then ask, as originally done by Barrow in the
context of black holes \cite{BARROW-MEMORY}, how the structure of the
astrophysical object is affected given that the asymptotic gravitational
coupling continues to evolve after the object forms.  Two possibilities
exist; either the star can adjust its structure in a quasi-stationary manner
to the asymptotic gravitational constant, or it might `remember' the strength
of gravity at the time it formed.  Barrow called this latter possibility {\em
gravitational memory}.  In the former case, stellar evolution is driven
entirely by gravitational effects, while in the latter case objects of the
same mass could differ in other physical properties, such as their radius.
Either possibility has fascinating consequences, which we have already
explored in Ref.~\cite{BOSON-MEMORY}.

However, which of the two scenarios is correct remains unknown, either for
black holes or boson stars.  Since boson star solutions are non-singular,
they appear to offer better prospects for determination the actual
behaviour.  Consequently, it is important at this stage to have a complete
description of boson star models at different eras of cosmic history, which
may be used later either as an initial condition for, or to compare with the
output of, a dynamical evolutionary code.

Recently, two other works have been presented concerning scalar--tensor
gravity effects on equilibrium boson stars.  In the first of them, Comer and
Shinkai \cite{COMER} studied zero and higher node configurations for the
Damour--Nordtvedt approach to scalar--tensor theories \cite{4-comer}.  They
also studied stability properties of boson stars both at the present time and
in the past.  They concluded that no stable boson stars exist before a
certain cosmic time, due to all possible configurations possessing a positive
binding energy.  This result appears surprising, as the boson stars should
have no particular awareness of the present value of the gravitational
coupling, and it would appear a great coincidence that the transition between
instability and stability should occur at a recent cosmic epoch.  In fact,
their result has already been questioned by Whinnett \cite{WHINNETT}, in a
detailed discussion of the meaning of the boson star mass in scalar--tensor
theories.  Our results also indicate that boson stars may 
form and be stable at any cosmic epoch.  Finally, the dynamical
formation of boson stars was analyzed in Ref.~\cite{SHINKAI2}, where a 
similar behaviour to that of general relativity was found.

In this paper, we aim to provide a comprehensive study of equilibrium
configurations of boson stars, emphasizing their characteristics, such as
mass and radii, at different moments of cosmic history.  We shall also study,
using catastrophe theory, their stability properties.  As seen in
Ref.~\cite{TORRES_BOSON}, the features of JBD and general scalar--tensor
boson stars do not differ much.  Hence, we shall concentrate only on JBD
boson stars, examining the dependence on $\omega$.  Finally, we shall test
whether the Brans--Dicke scalar can induce any change in the stability
properties even for extreme values of Newton's constant.

The organization of the rest of this work is as follows.  In the next section
we briefly introduce the formalism, following Ref.~\cite{TORRES_BOSON}.  The
following section will analyze some recently-proposed mass functions for JBD
boson stars, and justify our choice for this work.  We shall also
comment on the use of catastrophe theory in the study of stability
properties.  Finally, the results of our numerical simulations will be given 
in Sec.~V and our conclusions will be stated in Sec.~VI.

\section{Formalism}

First we derive the equations corresponding to a general 
scalar--tensor theory. The action for these generalized JBD
theories is
\begin{equation}
\label{action}
S = \int \frac{\sqrt{-g}}{16\pi} \, dx^4\left[ \phi R-
	\frac{\omega(\phi )}{\phi} \, \partial_\mu \phi \,
	\partial^\mu \phi + 16 \pi {\cal L}_{{\rm m}} \right] \,.
\end{equation}
Here $g_{\mu\nu}$ is the metric, $R$ the scalar curvature, $\phi$ the 
Brans--Dicke field, and ${\cal L}_{{\rm m}}$ the Lagrangian of the matter 
content of the system.

We take this  ${\cal L}_{{\rm m}}$ to be the Lagrangian density of 
a complex, massive, self-interacting scalar field $\psi$. 
This Lagrangian reads as:
\begin{equation}
{\cal L}_{{\rm m}} = -\frac{1}{2} g^{\mu \nu} \, \partial_\mu \psi^*
	\partial_\nu \psi -\frac{1}{2} m^2 |\psi|^2 
	-\frac{1}{4} \lambda |\psi|^4 \,.
\end{equation}
The $U(1)$ symmetry leads to conservation of boson number. Varying the 
action with respect to $g^{\mu\nu}$ and $\phi$ we obtain the field 
equations:
\begin{eqnarray}
\label{field0}
R_{\mu\nu}-\frac{1}{2} g_{\mu\nu}R & = & \frac{8\pi}{\phi}T_{\mu\nu}
	+\frac{\omega(\phi )}{\phi} \left( \phi_{,\mu} \phi_{,\nu}-
	\frac{1}{2} g_{\mu \nu }\phi^{,\alpha}\phi_{,\alpha}\right)
	\nonumber \\
 & & \quad + \frac{1}{\phi} \left( \phi_{,\mu;\nu} -g_{\mu\nu} \Box\phi
	\right) \,, \\
 \label{field00}
\Box \phi & = & \frac{1}{2\omega+3} \left[ 8\pi T-\frac{d\omega}{d\phi}
	\phi^{,\alpha}\phi_{,\alpha}\right] \,,
\end{eqnarray}
where $T_{\mu\nu}$ is the energy--momentum tensor for the
matter fields and $T$ its trace.
This energy--momentum tensor is given by
\begin{eqnarray}
\label{emt}
T_{\mu\nu} =\frac{1}{2} \left( \psi^*_{,\mu }\psi_{,\nu } +\psi_{,\mu }
	\psi^*_{,\nu }\right) -\frac{1}{2} g_{\mu\nu} ( g^{\alpha\beta}
	\psi^*_{,\alpha }\psi_{,\beta} + \nonumber \\ 
	m^2 |\psi|^2 + \frac{1}{2} \lambda |\psi|^4 ).
\end{eqnarray}
Commas and semicolons are 
derivatives and covariant derivatives, respectively. 
The covariant derivative of this tensor is null.
That may be proved either from the field equations, recalling the
Bianchi identities, or by intuitive arguments such as
the minimal coupling between the field $\phi$ and the matter 
fields. This implies, 
\begin{equation}
\label{fieldb}
\psi^{,\mu}_{~~;\mu}-m^2 \psi -\lambda |\psi|^2  \psi^*=0.
\end{equation}

We now introduce the background metric, corresponding to
a spherically-symmetric system which is the symmetry we impose
upon the star. Then
\begin{equation}
\label{metric}
ds^2=-B(r) \, dt^2 + A(r) \, dr^2 +r^2 d\Omega^2 \; . \label{METRIC}
\end{equation}
We also demand a spherically-symmetric form for the scalar field
describing the bosonic part and we adopt a form consistent with the
static metric,
\begin{equation}
\label{boson}
\psi(r,t)=\chi(r) \exp{[-i\varpi t]}.
\end{equation}
To write the equations of structure of the star, we use a rescaled radial 
coordinate, given by
\begin{equation}
\label{x}
x=mr \,.
\end{equation}
{}From now on, a prime will denote a derivative with respect to the variable
$x$. We also define dimensionless quantities by
\begin{equation}
\label{dimensionless}
\Omega=\frac{\varpi}{m}\; , \; \; \Phi=\frac{\phi}{m_{{\rm Pl}}^2} \; ,\; \; 
  \sigma=\sqrt{4\pi} \, \frac{\chi(r)}{m_{{\rm Pl}}} \; ,\; \;  
  \Lambda=\frac{\lambda}{4\pi} \, \frac{m_{{\rm  Pl}}^2}{m^2} \,,
\end{equation}
where $m_{{\rm Pl}} \equiv G_0^{-1/2}$ is the present Planck mass. Note that 
our dimensionless variables are defined with respect to our observed Planck 
mass, regardless of whether or not that corresponds to the Planck mass at 
that time.
Our 
observed gravitational coupling implies $\Phi = 1$.\footnote{There is 
actually a post-Newtonian correction to this of order $1/\omega$ 
\cite{12-boson}, which we 
shall not concern ourselves with.} In order to 
consider the total amount of mass of the star within a radius $x$ we change 
the function $A$ in the metric to its Schwarzschild form, 
\begin{equation}
\label{M}
A(x)=\left(1-\frac{2M(x)}{x\,\Phi(\infty)}\right)^{-1}.
\end{equation}
This expression {\em defines} $M(x)$.
The issue of mass definitions in JBD boson stars will
be examined more deeply in the following section.
Note that a factor $\Phi(\infty)$ appears in Eq.~(\ref{M}). This
is crucial to obtain the correct value of the mass, which from comparing 
this to the asymptotic form of the JBD--Schwarzschild solution is given by 
\begin{equation}
\label{schmass}
M_{{\rm star}}= M(\infty) \, \frac{m_{{\rm Pl}}^2}{m}\,,
\end{equation}
for a given value of $m$.\footnote{This corrects an error in Eq.~(10) of 
Ref.~\cite{BOSON-MEMORY}. That error was typographical only and did not 
affect any computations in that paper.}

With all these definitions, the equations of structure reduce to the 
following set:
\begin{eqnarray}                     
\label{field1}
\sigma^{\prime \prime} & + & \sigma^{\prime} \left( \frac{B^\prime}{2B} -
	\frac{A^\prime}{2A} + \frac{2}{x} \right) \nonumber \\
  && + A \left[ \left(\frac{\Omega^2}{B}-1 \right)\sigma - 
	\Lambda \sigma^3 \right]=0 \; , 
\end{eqnarray}
\begin{eqnarray}
\label{field2}
\Phi^{\prime \prime} & + & \Phi^{\prime} \left( \frac{B^\prime}{2B} -
	\frac{A^\prime}{2A} + \frac{2}{x} \right)+
	\frac{1}{2\omega+3}\frac{d\omega}{d\Phi} \Phi^{\prime 2}
	\nonumber \\
 && - \frac{2A}{2\omega+3} \left[ \left(
	\frac{\Omega^2}{B}-2 \right)\sigma^2 -\frac{\sigma^{\prime 2}}{A} - 
	\Lambda \sigma^4 \right] =0 \; , 
\end{eqnarray} 
\begin{eqnarray}
\label{field3}
\frac{B^{\prime}}{xB} & - & \frac{A}{x^2}\left( 1-\frac{1}{A} \right)=
	\frac{A}{\Phi} \left[ \left( \frac{\Omega^2}{B}-1 \right)
	\sigma^2 +\frac{\sigma^{\prime 2}}{A} - 
	\frac{\Lambda}{2} \sigma^4 \right] \nonumber \\
 && + \frac{\omega}{2}\left(\frac{\Phi^{\prime}}{\Phi}\right)^2
	+ \left( \frac{\Phi^{\prime \prime}}{\Phi}-
	\frac{1}{2}\frac{\Phi^{\prime}}{\Phi} \frac{A^\prime}{A} 
	\right) + \frac{1}{2\omega+3}\frac{d\omega}{d\Phi}
	\frac{\Phi^{\prime 2}}{\Phi} \nonumber \\
 &&  - \frac{A}{\Phi} \frac{2}{2\omega+3} \left[ 
	\left(\frac{\Omega^2}{B}-2 \right)\sigma^2
	-\frac{\sigma^{\prime 2}}{A} - \Lambda \sigma^4 \right] \; , 
\end{eqnarray}
\begin{eqnarray}
\label{field4}
\frac{2BM^\prime}{x^2 \Phi(\infty)} & = & \frac{B}{\Phi} \left[ \left(
	\frac{\Omega^2}{B}+1 \right)\sigma^2 +\frac{\sigma^{\prime 2}}{A}
	+ \frac{\Lambda}{2} \sigma^4 \right] \\
 & + & \frac{B}{\Phi} \frac{2}{2\omega+3} \left[
	\left(\frac{\Omega^2}{B}-2 \right)\sigma^2
	-\frac{\sigma^{\prime 2}}{A} - \Lambda \sigma^4 \right] \nonumber \\
 & + & \frac{\omega}{2}\frac{B}{A}\left( \frac{\Phi^{\prime}}{\Phi}\right)^2
 	-\frac{B}{A(2\omega+3)}\frac{d\omega}{d\Phi} 
	\frac{\Phi^{\prime 2}}{\Phi} - \frac{1}{2}
	\frac{\Phi^{\prime}}{\Phi} \frac{B^\prime}{A} \nonumber \,.
\end{eqnarray}
To solve these equations numerically, we use a fourth-order Runge--Kutta
method, for which details may be found in Ref.~\cite{TORRES_BOSON}.
In general relativity, the possible equilibrium solutions are entirely 
parametrized by the central value of the boson field, $\sigma(0)$. In JBD 
theory, one also needs to specify the asymptotic strength of the 
gravitational coupling, $\Phi_\infty$, or equivalently the value of $\Phi$ at 
the centre of the star.

The particle number, conserved due to the $U(1)$ symmetry of the $\psi$ 
field, is 
\begin{equation}
N_{{\rm star}}= \frac{m_{{\rm Pl}}^2}{m^2}  \Omega \int_0^\infty  \sigma^2 
	\sqrt{\frac AB} \, x^2 \, dx  \equiv \frac{m_{{\rm Pl}}^2}{m^2}
	N_\infty \,,
\end{equation}
where the last equality defines $N_\infty$. If the particles comprising the 
star were widely separated, their mass would be $mN_{{\rm star}}$. One can 
therefore define a binding energy, ${\rm BE}_{{\rm star}} = M_{{\rm star}} - 
m N_{{\rm star}}$, and a necessary, though not sufficient, condition for the 
star to be stable is that the binding energy be negative. Considerable care 
is however necessary in deciding how to define the mass which appears in this 
expression \cite{WHINNETT}, and we discuss this at length in the next 
Section.  It is normally convenient to consider a dimensionless binding 
energy, defined by
\begin{equation}
{\rm BE} = M(\infty) - m N_\infty \,.
\end{equation}

Finally, to get a feeling for the possible rate of variation of 
$\Phi$, we consider the solution corresponding to 
homogeneous matter-dominated cosmologies, which is \cite{Nariai,Gurevich}: 
\begin{equation}
\Phi(t) \propto t^{2/(4+3\omega)} \propto a^{1/(1+\omega)}\,.
\end{equation}
At the current limit $\omega =500$, the variation in $\Phi$ since 
matter--radiation equality at around $z_{{\rm eq}} = 24\,000\, \Omega_0 h^2$ 
is a couple of percent. During radiation domination $\Phi$, and hence $G$, 
is constant.

\section{Mass definitions}

The definition of mass in scalar--tensor theories is a subtle one, which has 
recently been examined in detail by Whinnett \cite{WHINNETT}.
When one leaves the security of general relativity, one first has to worry 
about which conformal frame one should work in, either the original Jordan 
frame as given in Eq.~(\ref{action}), or the Einstein frame obtained by 
carrying out a conformal transformation to make the gravitational sector 
match general relativity. Additionally, while in the Einstein frame all 
reasonable definitions coincide, in the Jordan frame they do not.

Whinnett studied three possible definitions.  He found huge differences for 
$\omega = -1$, but
the three definitions approach each other in the large $\omega$ limit, as one 
expects since they coincide in general relativity.  These
are to be compared with the {\it rest mass}, which is just the particle
number multiplied by the particle mass.  The definitions are:
\begin{itemize}
\item The Schwarzschild mass, given by
\begin{equation}
m(r)=4 \pi \int_0^r \rho r^2 dr \; ,
\end{equation}
where $\rho$ is defined as the right-hand side of
Einstein's timelike equation. This corresponds to the ADM mass in the Jordan 
frame. It is the commonly-used definition of mass and in the limit $r \to 
\infty$ coincides with $M_{{\rm star}}$ defined in Eq.~(\ref{schmass}).

\item The Keplerian mass, given by,
\begin{equation}
m_K(r)= r^2  \frac {B^\prime}{2} \; .
\end{equation}

\item The Tensorial mass, given by,
\begin{equation}
m_T (r)= r^3  \frac {B^\prime \phi + \phi^\prime B}
  {2  \phi r + r^2 \phi^\prime} \; .
\end{equation}

\end{itemize}

The last two definitions are orbital masses. A non-self-gravitating
test particle in a circular geodesic motion in the geometry of
Eq.~(\ref{METRIC}) moves with an angular velocity given by
\begin{equation}
\frac {d \varphi}{dt} = \sqrt { \frac{B^\prime}{2r} } \; ,
\end{equation}
as measured by an observer at infinity \cite{GRAVITATION}. Then, applying
Kepler's third law, the mass of the system can be obtained by making,
\begin{equation}
M(\infty)= \lim_{r \rightarrow \infty}
\left[ r^3 \left( \frac{d\varphi}{dt} \right)^2 \right] \; .
\end{equation}
So, the Keplerian mass is {\it Kepler's third law mass} in the Jordan frame,
whereas the Tensorial mass is {\it Kepler's third law mass} in the Einstein
frame. In the Einstein frame all mass definitions 
coincide, so the Tensorial mass is also the Einstein frame ADM mass.

These definitions differ impressively for the $\omega =-1$ case, and, in
general, for low values of $\omega$ \cite{WHINNETT}.  Then, of course, it
becomes very important to have a correct description of the stellar mass,
because it will decide stability properties and binding energy behaviour.  
For the case $\omega=-1$, the Keplerian mass would lead to positive binding
energy for all values of central density, suggesting that every solution is
generically unstable.  The Schwarzschild mass would instead lead to negative
binding energies for every value of central density, suggesting that every
solution is potentially stable, even for large values of
$\sigma(0)$.\footnote{In fact, for small values of central density the
Schwarzschild mass becomes negative for low $\omega$.  This might indicate
that a classical wormhole can form, in much in the same way as the solution
presented in Ref.~\cite{TORRES_BDWH}.}  This leads one to feel that neither
of these two masses is likely to be the correct one to use in the binding
energy calculation.  Further, it is the Tensorial mass which peaks (as a
function of central density) at the same location as the rest mass, an
important property in the general relativity case \cite{HARRISON} which is 
crucial in
permitting the application of catastrophe theory to analyze the stability
properties.  This property presumably originates from the Tensorial mass
being the Einstein frame ADM mass, though we have no mathematical proof at
present.  There is therefore a strong case \cite{WHINNETT} towards the 
adoption of the Tensorial mass as the real mass of the star, especially 
for the strong field cases of low $\omega$ values.

For the simulations we analyze in this work, we have computed both the
Schwarzschild and the Tensorial mass. As expected, we find that for large 
$\omega$ values, which are the ones in which we are interested,
the difference is negligible; every graph we plot is unchanged if we replace 
the Schwarzschild mass by the Tensorial mass. Hence, for reasons of
numerical simplicity we actually compute the Schwarzschild mass, as it is 
directly obtained from the set of differential field equations.

\section{Stability analysis using catastrophe theory}

Catastrophe theory provides a very direct route to the stability properties
of boson stars \cite{KUS}.  The technique was described in a review of the
stability of solitons \cite{KUS0}, in which it was shown that the
identification of conserved quantities of a physical system is sufficient for
the determination of stable und unstable solitons.  In the case of boson
stars, we are dealing with nontopological solitons which are characterized by
mass and particle number, the only conserved quantities of this theoretical
model.  For every central value of the scalar field, there is a unique value
for the mass and particle number.  By drawing the conserved quantities
against each other, the so-called {\em bifurcation diagram} is created.  If
{\em cusps} are present in this diagram, one can immediately read off the
stable and unstable states.  Starting with small central densities where mass
and particle number is also small, one assumes that these stars are stable
(against small radial perturbations).  If, as the central density is
increased, one meets a cusp, the stability of the boson star changes from
stable to unstable if the following states --- the branch as a whole --- have
higher mass.  This method is applied again at every succeeding cusp.  Should
it be that at some cusp the masses beyond the cusp are smaller, then the
state changes from unstable to stable.  The reason behind this method is that
the cusp is a projection of a saddle point of a Whitney surface \cite{KUS}.
The curve leading to the cusp consists of projections of fold points; fold
points and cusps are the singularities of the Whitney surface, just the
points recognizable in the bifurcation diagram.  The fold points are the
projection of maxima and minima of Whitney's surface; maxima determine
unstable solutions while minima govern stability.

\begin{figure}[t]
\centering 
\leavevmode\epsfysize=12cm \epsfbox{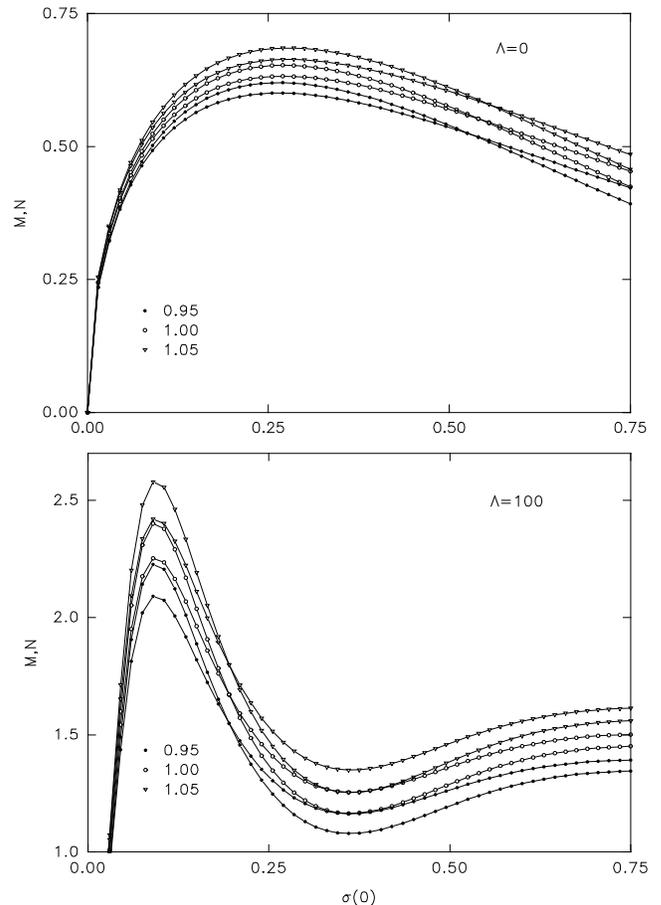}\\ 
\caption[fig1]{\label{fig1} Boson star equilibrium configurations for 
different values of $\Phi(\infty)$ and self-interaction. Of the two lines 
for each case, the one with the higher peak is the particle number. 
$\sigma(0)$ is in the interval $(0,0.75)$ and there are 50 models per curve.}
\end{figure} 

The method of catastrophe theory has also been applied in the context of
neutron stars \cite{KUS1}, Einstein--Yang--Mills black holes \cite{MAEDA},
and inflationary theory \cite{KUS2}.  More recently, it has been introduced
for neutron and boson stars in scalar--tensor theories, and in particular in
JBD theory \cite{COMER,HARADA}.  {}From these theories, one can learn how the
properties of static solutions change if the asymptotic value of $\Phi$
changes.  In the following, we show that there is no stability change at all
within a JBD theory as $\Phi_\infty$ is changed; if a star is stable, then it
is for every value of $G$.  The binding energy can change its sign for stars
which are unstable, but not for stable ones.

\begin{figure}[t]
\centering 
\leavevmode\epsfysize=12cm \epsfbox{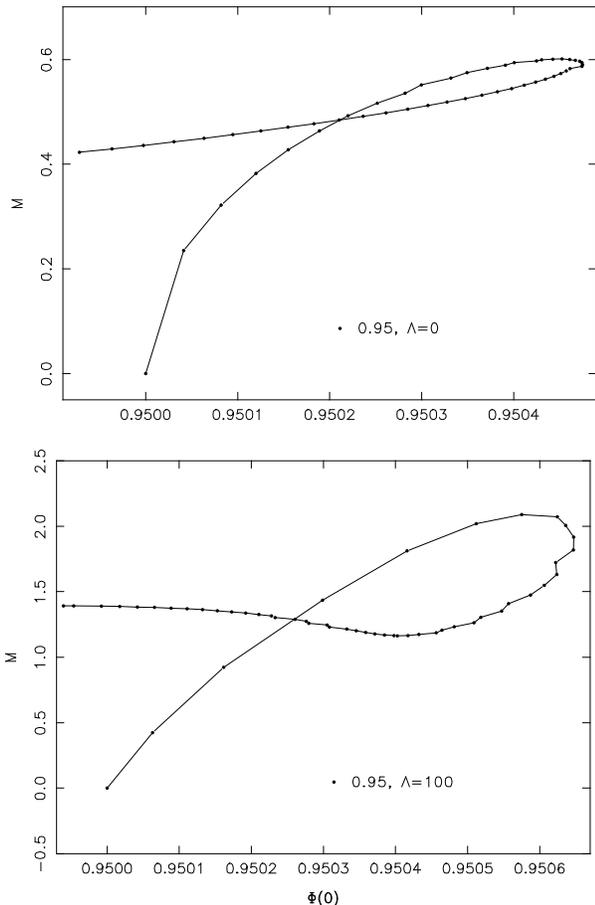}\\ 
\caption[fig2]{\label{fig2} Typical curves for boson star masses as a 
function of $\Phi(0)$, with $\Phi(\infty) = 0.95$. Note the narrow $x$-axis 
range.}
\end{figure} 

\section{Numerical Results}

\begin{figure}[t]
\centering 
\leavevmode\epsfysize=12cm \epsfbox{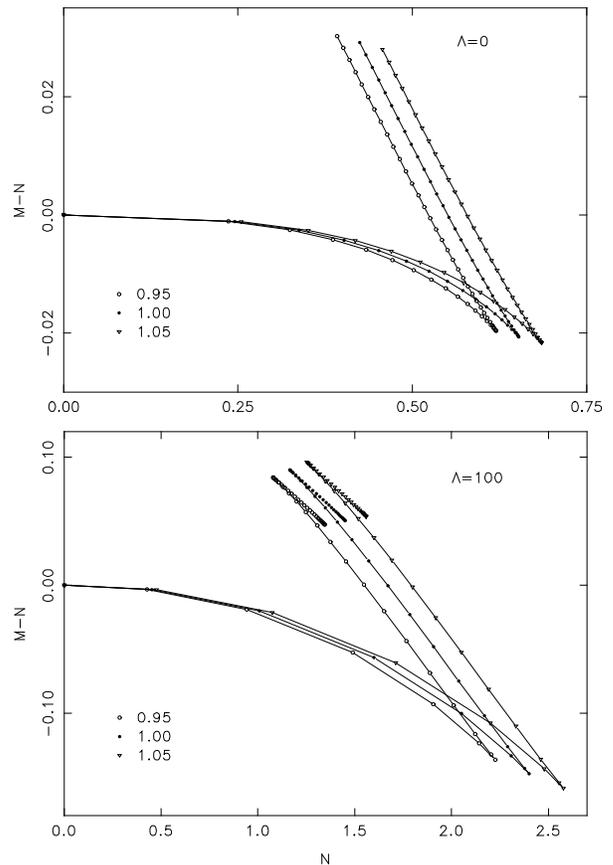}\\ 
\caption[fig3]{\label{fig3} The boson star binding energy as a function of 
the number of particles. The figure depict several curves for different 
$\Phi(\infty)$
and $\Lambda$
values, the coupling is $\omega=400$.} 
\end{figure} 

First, we plot the equilibrium configuration diagrams for different values of
the central density and asymptotic gravitational constant.  In Fig.~1a
($\Lambda=0$), we have 50 models with central density in the interval
$(0,0.75)$, with no self-interaction.  Fig.~1b shows the same, but with
$\Lambda=100$.  We recognize that at fixed central density $\sigma(0)$, the
mass and particle number increase from earlier times ($\Phi(\infty)=0.95$) to
later times ($\Phi(\infty)=1.05$).  If we draw the mass against the
central value of the JBD field $\Phi(0)$, we find a loop, see
Fig.~\ref{fig2}.  The curve starts at the flat spacetime solution ($\Phi=$
constant everywhere and zero mass), reaches the maximum at the same value of
central scalar field as it reached the maximum of Fig.~1, cf.~\cite{COMER},
makes a turn, and eventually reaches smaller $\Phi(0)$ values.  Stable stars
are characterized by $\Phi(0)>\Phi(\infty)$, i.e.~$G(0)<G(\infty)$ where $G$
is a function of $r$.  Unstable stars can have $G(0)$ greater than or less
than $G(\infty)$.  There are two solutions for $\Phi(0)=\Phi(\infty)$:
first, the flat spacetime solution and, secondly, an unstable boson star.
The same characteristic curve is to be found for different values of the
asymptotic $G$.

Figs.~\ref{fig1} and \ref{fig2} give us, in form of
$(\sigma(0),\Phi(0))$, the complete information about the initial
characteristics of a boson star at a certain `time', characterized by the 
constant $\Phi(\infty)$.

\begin{figure}[t]
\centering 
\leavevmode\epsfysize=9cm \epsfbox{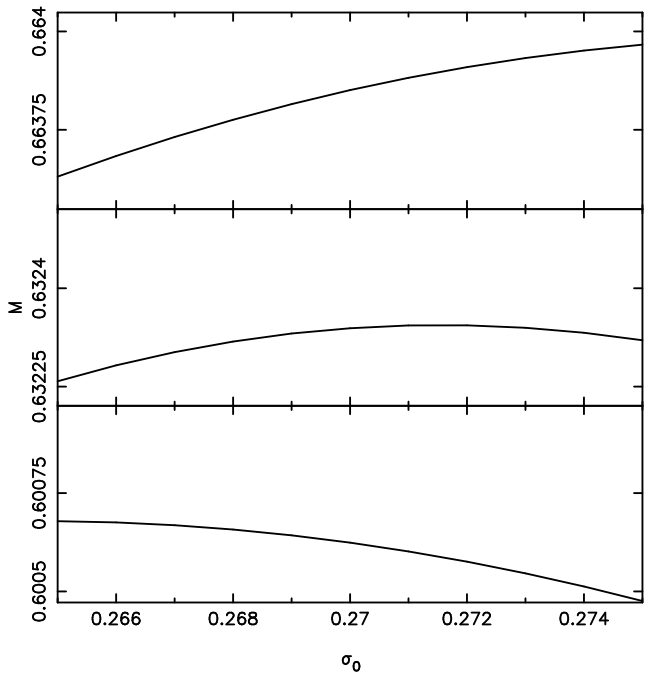}\\  
\caption[fig-new]{\label{fig-new} Boson star masses for a reduced interval 
of values of $\sigma(0)$ around the cusp at three different `times'
$\Phi(\infty)=1.05$ (top), $\Phi(\infty)=1.0$ (middle), and
$\Phi(\infty)=0.95$ (bottom) [$\omega =400$].}
\end{figure}

For the investigation of stability, rather than the bifurcation diagram
$(M,N)$ we use the analogous figure of binding energy against the particle
number, Fig.~\ref{fig3}.  It shows us that the stars with small central
densities have negative binding energies.  In Fig.~3b ($\Lambda=100$) we see
two cusps:  the first one corresponds to the maximum of Fig.~1 and the second
to the minimum.  For $\Lambda = 0$ we did not go to high enough central
densities to see the second cusp.  The first cusp has negative binding energy
and the other has positive binding energy.  Stars with central densities from
zero to the first cusp belong to projections of minima within a Whitney
surface, i.e.~they are stable.  Beyond the cusp, one radial perturbation mode
is becoming unstable, and at the second cusp a second mode becomes unstable.

To study the influence of the changing gravitational coupling on the exact
position of the cusp we made a high resolution study around the position of
the cusp for the $\Phi(\infty)=1$ model.  As can be seen from from Fig.~1, it 
is at $\sigma(0) \simeq 0.27$.  We then did simulations in the 
interval $\sigma(0) \in [0.265, 0.275]$, with eleven
models in that range, shown in Fig.~4.  For $\Phi(\infty)=1.05$ the mass and 
the number
of particles as a function of $\sigma(0)$ are increasing functions.  That
means all the models are in the stable branch. For $\Phi(\infty)=0.95$ 
instead, mass and number of particles are decreasing functions, locating
all models in the unstable branch.  In the case of our present gravitational
coupling, the cusp appears within the interval.  Thus, going from the future
to the past ($\Phi(\infty)=1.05$ to $\Phi(\infty)=0.95$) models with a given 
central density move from the
stable to the unstable branch. This agrees with the analytic prediction that 
in the general relativity limit one should find $\sigma^2_{{\rm max}} \propto 
\Phi(\infty)$. The movement of the cusp is much the same as Comer and Shinkai 
reported in \cite{COMER}, except for one important point.  They found no cusp
at all for times well before the present, meaning that they did not find {\it 
any}
stable star in the past.  On the contrary, we have found that the cusp moves
backwards in $\sigma(0)$, but it is still there, see Fig.~3.  We believe
that their conclusion derives from the use of a wrong mass definition.

\begin{figure}[t]
\centering 
\leavevmode\epsfysize=11cm \epsfbox{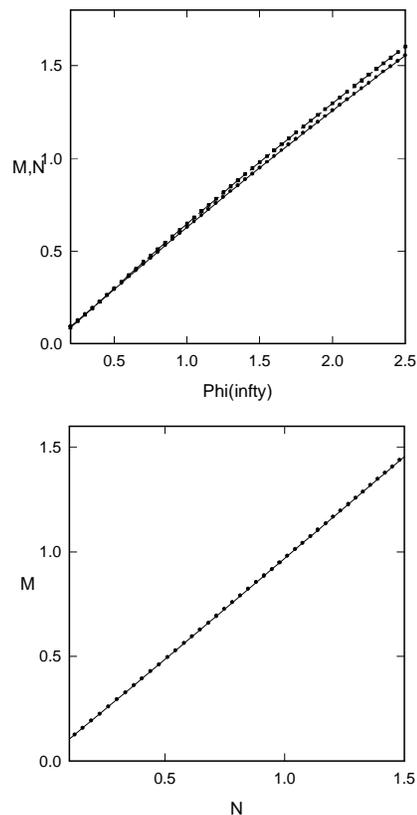}\\
\caption[fig4]{\label{fig4} The behaviour of masses and particle numbers for 
extreme values of the gravitational asymptotic constant.
The model taken has $\Lambda=0$, $\sigma(0)=0.1$ and $\omega=400$.}
\end{figure} 

In addition, we have  calculated
solutions with constant central scalar field values at different `times'
$\Phi(\infty)$, also taking into account very small values of $\Phi(\infty)$
which are unphysical, see Fig.~\ref{fig4}.  This figure represents a
bifurcation diagram with respect to $\Phi(\infty)$.  It is evident that no
cusp is present, so no stability change occurs.  

\begin{figure}[t]
\centering 
\leavevmode\epsfysize=13cm \epsfbox{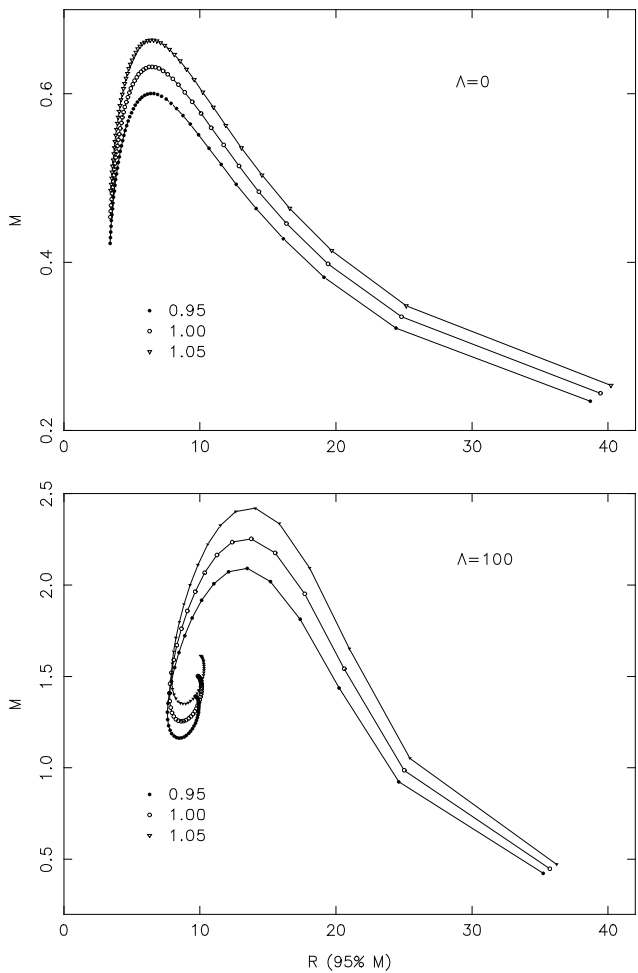}\\  
\caption[fig5]{\label{fig5} The radius of equilibrium boson star 
configurations for different values of the 
self-interaction and central density in the range $(0,0.75)$.}
\end{figure} 

Because a boson star has no clearly-defined surface, but rather an infinite
exponentially-decreasing atmosphere, several radius definitions are in use.
We apply here the common one, the radius which encloses 95\% of the total
mass.  Fig.~\ref{fig5} represents the mass against the radius.  The diagram
shows that solutions with small central densities have large radii.  Then,
with growing central densities, the mass increases while the radius
decreases.  The maximum in this diagram is the most centrally-dense stable 
star solution. The radius of the maximal
mass boson star remains roughly 
the same, but the mass corresponding to a given
central density grows with time, producing a denser star.  The increase of
the self-interaction constant $\Lambda $ gives larger radii as one expects
from a repulsive force.  Compare this with similar results for neutron
stars (Figure 7 in Ref.~\cite{HARRISON}) and for general relativistic boson 
stars (Figure 3 in Ref.~\cite{KUS1}).

\begin{figure}[t]
\centering 
\leavevmode\epsfysize=6.5cm \epsfbox{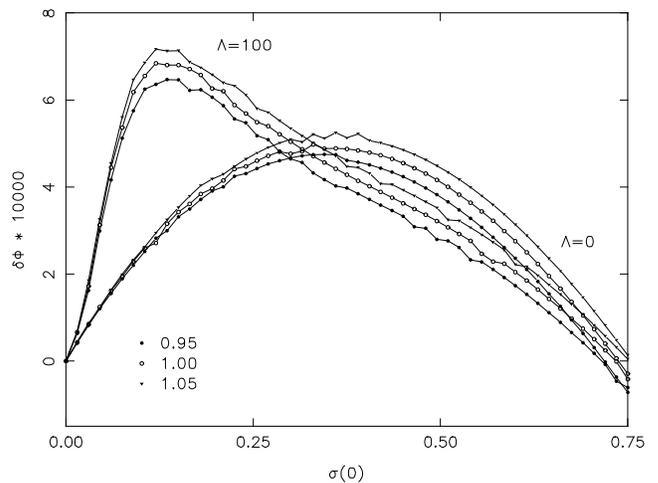}\\  
\caption[fig6]{\label{fig6} The behaviour of the difference between the 
central and the asymptotic value of the Brans--Dicke scalar as a function of 
$\sigma(0)$ for different values of the effective gravitational constant. 
Note the highly expanded $y$-axis.}
\end{figure}

So far, we have recognized that the boson stars are denser the larger 
$\Phi(\infty)$ is. The reason can be understood as a
deeper gravitational potential, expressed by an increase in the difference 
of $\Phi(0)$ and $\Phi(\infty)$, see Fig.~\ref{fig6}.

\begin{figure}[t]
\centering 
\leavevmode\epsfysize=9.9cm \epsfbox{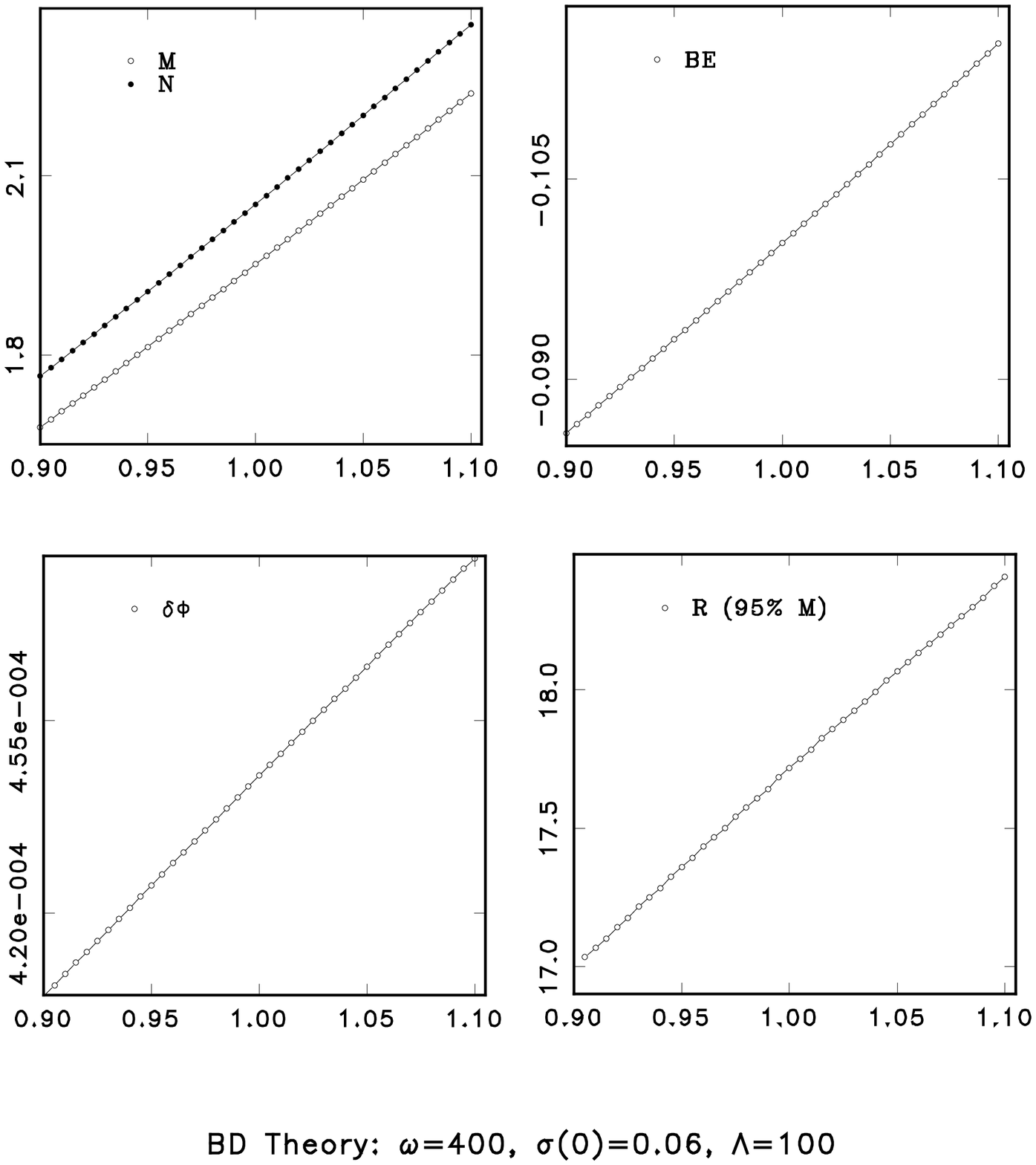}\\  
\caption[fig7]{\label{fig7} Boson star features for a given central density, 
as a function of $\Phi_\infty$.}
\end{figure}

For a fixed value of $\sigma(0)$, the behaviour of the binding energy,
the radius, and $\delta \Phi$ (the difference between the central and the 
asymptotic value of the Brans--Dicke scalar) are all plotted in 
Fig.~\ref{fig7}.

\begin{figure}[t]
\centering 
\leavevmode\epsfysize=10cm \epsfbox{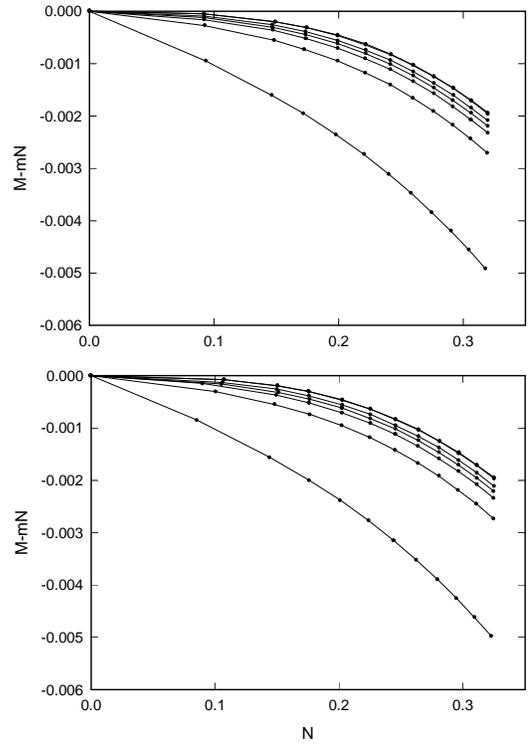}\\ 
\caption[fig8]{\label{fig8} Boson star binding energy as a function of the 
number of particles, for different values of $\omega$ ranging 
from 50 to 50000. The upper panel shows models with $\Phi(\infty)=0.98$, the 
lower one, with $\Phi(\infty) =1$. The value of $\sigma(0)$ lies in the 
range $(0,0.3)$.}
\end{figure}

In Fig.~\ref{fig8}, we show the dependence of equilibrium configurations on 
$\omega$.  To do
this we plot the binding energy behaviour in the interval $\sigma(0) \in
(0,0.3)$ for two values in the asymptotic effective gravitational constant.
The value of $\omega $ is in the range (50, 50000).  The upper curves in both
diagrams correspond to $\omega =10000$ and $50000$ and match each
other exactly. The upper panel shows models with $\Phi(\infty)=0.98$, while
the other has our observed gravitational strength.  This shows that,
when $\omega$ tends to infinity, a general relativity like solution --- with
a different value for Newton's constant --- is obtained.  Recall that even
with the strong limit on $\omega$ valid today, we could have an evolving
$\omega(\phi)$ which is much smaller in the past, and so small values of the 
coupling parameter may also be meaningful.

\section{Concluding remarks}

In this paper, we have thoroughly analyzed static boson star configurations
in the framework of the Jordan--Brans--Dicke theory of gravitation.  We
studied their equilibrium and stability properties in the present as well as
for other cosmic times, in the past or in the future.  Stable boson stars may
exist at any epoch, with stability depending on the value of central density.
Together with this, a number of new physical features have been displayed
concerning the radius--mass relation, the behaviour of the difference between
the central and asymptotic value of $\Phi$, the dependence on the structure
upon the coupling parameter and other properties.  This configurations can be
used either to compare with the output of a numerical evolution code, or as
the input into one.  We expect that such a study will shed light on which
scenario of the gravitational memory phenomenon might occur in practice.
Whichever it might be, it is very likely that the same phenomena could also
occur for fermionic stars, such as white dwarfs.  In this sense, the results
obtained in this paper can be regarded as of a general nature.  Astrophysics
should be unambiguously sensitive to the underlying theory of gravity,
especially on cosmological times scales.  It is in this framework, perhaps,
where a crucial test of gravity could arise.

\acknowledgments

D.F.T. was supported by a British Council Fellowship (Chevening Scholar) 
Fundaci\'on Antrochas and CONICET, F.E.S.~by a
European Union Marie Curie TMR fellowship and A.R.L.~by the Royal Society.
We are indebted to Andrew Whinnett for a series of valuable discussions.

\end{document}